\documentstyle[12pt]{article}
\begin{document}
\hfill SIM/MATH/96-01

%\hfill June 1996
\vspace*{2cm}
\begin{center}
{\Large\bf 
TOPOLOGY, QUANTUM GRAVITY\\
AND\\
PARTICLE PHYSICS}\\[1.5cm]
{\large Andrew Toon\footnote{E-mail: andrewtoon@sim.ac.sg}\\
Singapore Institute of Management\\
Open University Degree Programmes\\
Mathematics Department\\
535A Clementi Road\\
Singapore 599490}
\end{center}
\begin{abstract}
It is argued that quantum gravity has an interpretation as a topological
quantum field theory provided a certain constraint from the path integral
measure is respected. The constraint forces us to couple gauge
and matter fields to gravity for space - time dimensions different from 3.
We then discuss possible models which may be relevant to our universe.
\end{abstract}
\vfill
%\noindent PACS: 02.30+g, 02.40. + m
\newpage

\section{Introduction}

Why do we live in a 4 dimensional space - time? Why does the standard model
have its particle content of 3 generations and why is its gauge group
$SU(3)\times SU(2)\times U(1)$? These are just a few fundamental questions
as yet unanswered by our current favorite theories.

String theory has for many years promised to answer these types of questions.
For example, the development of the heterotic string theory suggested that,
at the Planck scale, the number of space - time dimensions is 10 with an
$E_{8}\times E_{8}$ gauge group together with some particle content. However,
the theory does not tell us how to compactify, if at all, to 4 space - time
dimensions and how the gauge group $E_{8}\times E_{8}$ breaks down towards
the standard model around the electroweak scale. The development of various
compactification schemes, as well as the construction of intrinsically four
dimensional models, soon lead to a vast number of string theories which
would seem to ruin string theories initial uniqueness as far as anomaly
cancellation is concerned. However, it is hoped that most of these theories
are simply different phases of the same theory as exemplified by the program
of mirror symmetry.

What is needed is some kind of ^^ ^^ renormalisation group equation" for
the number of space - time dimensions, gauge group etc. which tells us how
these quantities vary with energy scale. This of course being a very
difficult question whose answer must await a non - perturbative approach to
string theory.

String theory is not the only possible answer to our fundamental questions.
Theories based on conventional field theories may still have a chance of
answering these deep issues but these kind of theories, admittedly, must be of
a new kind and draw upon new symmetries to come up with acceptable answers.

The approach we have in mind is topological field theories \cite{Bham,Thom}.
These theories
first made their impact by the way they reproduce topological invariants of
certain smooth manifolds \cite{Wit1}. From a physical point of view it is
hoped that
these theories may have something to do with quantum gravity. That is, one
imagines the initial phase of our universe to be in an unbroken phase where
there is no space - time metric. The universe we see would then correspond to
the so called broken phase where, via some mechanism, the initial unbroken
phase develops some metric dependence thereby generating a space - time
structure where physics can operate.

The model example of this is Witten's interpretation of 2 + 1 quantum gravity
as a Chern - Simons gauge theory with gauge group $ISO(2, 1)$, $SO(3, 1)$ or
$SO(2, 2)$ depending on whether the cosmological constant is zero, positive
or negative \cite{Wit2,Wit3}. This theory has two natural phases depending
on whether a
certain operator has zero modes or not and it is argued that each of these
phases corresponds to the unbroken and broken phases of gravity. However,
the theory has several draw backs as far as physics is concerned. There is a
deep problem with incorporating second quantised matter fields which in some
way must only manifest themselves in the broken phase but melt away in the
unbroken phase \cite{Wit3}. It also, of course, misses the physically
interesting ball park of four dimensional space - time.

There do exist topological field theories in four dimensions but at present
they seem to have nothing to do with gravity \cite{Horo}. This being our biggest
obstacle before we even begin to worry on how to incorporate second quantised
mater fields etc. into the theory. However, there exits topological field
theories in five dimensions which reproduces classical gravity in four
dimensions when appropriate boundary conditions are imposed on the five
dimensional topological field theory \cite{Brooks,Banados}.

In trying to make contact between physics and topological field theories,
most approaches start from the top down. That is, one starts off with a
topological field theory and attempt to make contact with geometry and hence
physics via some symmetry breaking mechanism. This of course is a familiar
approach but it is extremely difficult to see how the enormous symmetries
involved break down in a desirable way. Perhaps a better approach is to start
at the bottom and work up. That is, start off with conventional ideas of
quantum gravity and attempt to make contact with topological field theories.

An approach along these lines was attempted in \cite{Toon1}. Here, it was
argued that
quantum Einstein gravity, when coupled to appropriate fields, has an
interpretation as a topological field theory. The observations made here were
very interesting in that a powerful constraint was discovered which involved
the number of space - time dimensions, the dimensions of possible gauge
groups together with the possible dimensions of the representations that any
matter fields could transform under. In some sense, this constraint tells us
how the matter fields of the theory varies as the number of space - time
dimensions varies which in turn we imagine being related to the energy
scale of the theory.

The approach in \cite{Toon1} was in the second order formalism. We now wish to
perform a similar analysis in the first order formalism since in this
approach the theory is more like a gauge theory and we wish to include
fermionic matter.

\section{Topological field theories}

In this section we review the relevant facts of topological field theories
in order to discover what is needed in the case of quantum gravity for it to
have an interpretation as a topological field theory. In particular, we will
be interested in Schwarz type topological field theories where the crucial
point being
in the gauge fixing. It is this procedure which really tells us what we mean
by a Schwarz type topological field theory.

Consider then an $n$ - dimensional smooth manifold $M_{n}$ which we regard as
an $n$ - dimensional space - time. We assume we can construct some
classical action $S$ on $M_{n}$ which is general co-ordinate invariant
together with some gauge symmetry. That is, in our case, $S$ does not contain
any
space - time metric of $M_{n}$. To evaluate, for example, the partition
function corresponding to $S$ we must gauge fix. The important point here is
in order to fully gauge fix the theory we must choose some metric
$g_{ij}$ on $M_{n}$. The fully gauged fixed quantum action $S_{q}$
then takes the form\footnote{We will ignore such issues as Gribov ambiguities
in this paper.}:
\begin{equation}
S_{q}[\Phi,g_{ij}]=S[\Phi_{r}]+\delta_{Q}V[\Phi_{r},g_{ij}].
\end{equation}
Here $\Phi_{r}$ $(r = 1, 2, ...)$ are the fields of the theory including
matter, gauge, ghost and auxiliary fields etc. The second term on the
right hand side is the ghost plus gauge fixing term associated with all the
gauge invariance of $S[\Phi_{r}]$. $\delta_{Q}$ stands for the related
nilpotent BRS - transformation  and thus the entire gauge fixing plus ghost
term is written as a BRST variation of some functional $V[\Phi_{r},g_{ij}]$
The partition function is thus given by:
\begin{equation}
Z(M_{n},g_{ij})=\int D[\Phi]\exp iS_{q}[\Phi_{r},g_{ij}].
\end{equation}
What we mean by topological invariance is, for example, the partition
function $Z(M_{n},g_{ij})$ only depends on the gauge fixing metric $g_{ij}$
topologically:
\begin{equation}
\frac{\delta Z(M_{n},g_{ij})}{\delta g^{ij}}=\int D[\Phi]\exp(iS_{q})\frac
{\delta
(\delta_{Q}V)}{\delta g^{ij}}=\int D[\Phi]\exp(iS_{q})\delta_{Q}(\frac{\delta V}
{\delta g^{ij}})=0,
\end{equation}
by BRST invariance \cite{Thom}. That is, a small variation in the metric
$g_{ij}$  will not change the partition function $Z(M_{n},g_{ij})$.

What is crucial in the above observation
is the path integral measure $D[\Phi]$ of equation (2) be independent
of the gauge fixing metric $g_{ij}$. This, in general, will not be the case
as we now discuss.

Since we are essentially considering a field theory on a curved manifold
with metric $g_{ij}$,
the partition function for the theory is not the naive one of equation (2)
but is given by:
\begin{equation}
\tilde{Z}(M_{n},g_{ij})=\int\tilde{D}[\Phi]\exp iS_{q}[\Phi_{r},g_{ij}],
\end{equation}
where $\tilde{D}[\Phi]$ stands for an appropriate general co-ordinate
invariant measure over the set of fields $\Phi_{r}$. Fujikawa \cite{Fuji} has
proposed the following choice of the measure for being co-ordinate invariant:
\begin{equation}
\tilde{D}[\Phi]=\prod_{x}[\prod_{r}d\tilde{\Phi}_{r}(x)],
\end{equation}
where for any given field component $\Phi_{r}$:
\begin{equation}
\tilde{\Phi}_{r}(x)=g^{\alpha_{r}}(x)\Phi_{r}(x),
\end{equation}
where $g(x)=\det g_{ij}$. The $\alpha_{r}$ are constants which depend upon
the tensor nature of the field as well as the number of space - time
dimensions. Specifically we have:
\begin{equation}
\alpha_{r}=\frac{n-2m}{4n},
\end{equation}
for each component of a covariant tensor of rank $m$ in $n$ - dimensions. It
is now clear that:
\begin{equation}
\prod_{x}d\tilde{\Phi}(x)=\prod_{x}d[g^{\alpha_{r}}(x)\Phi_{r}(x)]=\prod_{x}
[g^{\alpha_{r}\sigma_{r}}(x)d\Phi_{r}(x)],
\end{equation}
where the signature $\sigma_{r}$ is + $(-1)$ for commuting (ant - commuting)
fields. Thus, the general covariant Fujikawa measure becomes:
\begin{equation}
\tilde{D}[\Phi]=\prod_{x}[g(x)]^{K}[\prod_{r,y}d\Phi_{r}(y)],
\end{equation}
where:
\begin{equation}
K=\sum_{r}\sigma_{r}\alpha_{r},
\end{equation}
is an index which measures the $g$ - metric dependence of the path integral
measure. The partition function (4) thus becomes:
\begin{equation}
\tilde{Z}(M_{n},g_{ij})=\prod_{x}[g(x)]^{K}Z(M_{n},g_{ij}),
\end{equation}
where $Z(M_{n},g_{ij})$ is the partition function using the naive path
integral measure $D[\Phi]=\prod_{x,r}[d\Phi_{r}(x)]$. Thus, for topological
invariance to be
preserved at the quantum level with respect to the metric $g_{ij}$, we
require \cite{KR}:
\begin{equation}
K=\sum_{r}\sigma_{r}\alpha_{r}=0.
\end{equation}
The observations made in this section are the basic ideas regarding quantum
gravity as a topological field theory. We wish to write down a classical
action for quantum gravity in terms of vielbein and spin connection fields.
In order to quantise the theory we must gauge fix due to
diffeomorphism invariance. To gauge fix we pick some gauge fixing
metric $g_{ij}$ on
$M_{n}$ (unrelated to the vielbein fields) and argue that its corresponding
partition function does not depend upon the choice of $g_{ij}$. That is, we
may formally view quantum gravity as a topological quantum field theory
with respect to
the gauge fixing metric $g_{ij}$ provided the constraint (12) is satisfied. 

\section{Gravity as a gauge theory}
In this section, we wish to discuss gravity as a gauge theory in $n$
dimensional space - time. When $n=3$, gravity with a zero cosmological
constant has a natural interpretation as a Chern - Simons gauge theory with
gauge group $ISO(2,1)$ \cite{Wit2,Wit3}. However, in $n\neq 3$, these ideas
cannot
be generalised due to the fact that we cannot construct a non - degenerate
bilinear form on the Lie algebra of $ISO(n-1,1)$. Let us then discuss in some
detail in what sense we may construct a gauge theory of gravity in a $n$ -
dimensional space - time ($n>2$) with zero cosmological constant.

Gravity is most naturally thought of as a gauge theory associated with
rotations and translations of Minkowski space, that is the Poincar$\acute{e}$
group. The Poincar$\acute{e}$ group  in $n$ space - time dimensions
contains $n$  generators $P_{a}$
($a=0,1,2,..., n-1$) associated with space - time translations,
and $n(n-1)/2$ generators
$M_{ab}$ associated with Lorentzian rotations. The algebra of these generators
satisfies:
$$[P_{a},P_{b}]=0,$$
\begin{equation}
[M_{ab},P_{c}]=\eta_{bc}P_{a}-\eta_{ac}P_{b},
\end{equation}
$$[M_{ab},M_{cd}]=\eta_{ac}M_{bd}+\eta_{bd}M_{ac}-\eta_{bc}M_{ad}-\eta_{ad}
M_{bc},$$
where $\eta_{ab}$ is the Lorentz metric in $n$ dimensional space - time.

In this article, we consider a gauge theory of gravity to be a gauge theory
corresponding to the Poincar$\acute{e}$ algebra (13). That is, we regard the
Poincar$\acute{e}$ group as an internal symmetry. As a step in this
direction, let us define the covariant derivative:
\begin{equation}
D_{i}=\partial_{i}+e_{i}^{a}P_{a}+\frac{1}{2}\omega_{i}^{ab}M_{ab},
\end{equation}
where $e_{i}^{a}$ are the gauge fields associated with translations (the
vielbeins) with $i$ = 0, 1, 2, ..., $n-1$ representing curved space - time
indices
and $\omega_{i}^{ab}=-\omega_{i}^{ba}$ are the gauge fields associated with
Lorentzian rotations (the spin connections). We may now immediately calculate:
\begin{equation}
[D_{i},D_{j}]=R_{ij}^{a}P_{a}+\frac{1}{2}R_{ij}^{ab}M_{ab},
\end{equation}
where:
\begin{equation}
R_{ij}^{a}=\partial_{i}e_{j}^{a}-\partial_{j}e_{i}^{a}+e_{ib}\omega_{j}^{ab}
-e_{jb}\omega_{i}^{ab},
\end{equation}
and:
\begin{equation}
R_{ij}^{ab}=\partial_{i}\omega_{j}^{ab}-\partial_{j}\omega_{i}^{ab}+\omega
_{ic}^{a}\omega_{j}^{cb}-\omega_{jc}^{a}\omega_{i}^{cb}.
\end{equation}
An infinitesimal gauge parameter can be written
as $u=\rho^{a}P_{a}+(1/2)\tau^{ab}M_{ab}$ where $\rho^{a}$ and $\tau^{ab}$ are
infinitesimal parameters. Since, under a general gauge transformation, the
covariant derivative transforms as $D_{i}'=hD_{i}h^{-1}$, where $h=\exp u$
in our case, we deduce the
following transformation rules for the vielbein and spin - connection fields:
\begin{equation}
\delta e_{i}^{a}=-\partial_{i}\rho^{a}-\omega_{i\;\; b}^{a}\rho^{b}+\tau^{ab}
e_{ib},
\end{equation}
\begin{equation}
\delta\omega_{i}^{ab}=-\partial_{i}\tau^{ab}-\omega_{ci}^{\;\; b}\tau^{ac}+
\omega^{a}_{i\;\; c}\tau^{cb}.
\end{equation}
We now discuss in what sense we consider gravity as a gauge theory.
The most obvious requirement is to construct some action, involving the fields
$e_{i}^{a}$ and $\omega_{i}^{ab}$, which is invariant under the transformations
(18) and (19) and reproduces ordinary classical general relativity.
One way to construct the required action is to demand what one normally
considers to be the space - time metric, $g_{ij}=\eta_{ab}e_{i}^{a}e_{j}^{b}$,
to have the correct transformation properties under diffeomorhisms. This
requirement forces us to demand $e_{i}^{a}$ be invertible and
$R_{ij}^{a}$ of equation (16) to vanish. Thus, we require:
\begin{equation}
R_{ij}^{a}=\partial_{i}e_{j}^{a}-\partial_{j}e_{i}^{a}+e_{ib}\omega_{j}^{ab}
-e_{jb}\omega_{i}^{ab}=0
\end{equation}
to follow as a constraint from the action. An action with the required
properties is:
\begin{equation}
S=\int e^{a}\wedge e^{b}...\wedge R^{cd}(\omega)\epsilon_{ab...cd}=\int
e_{i}^{a}e_{j}^{b}...R^{cd}_{kl}\epsilon^{ij...kl}\epsilon_{ab...cd},
\end{equation}
where there are $n-2$ vielbeins $e$ and $R(\omega)=d\omega+\omega\wedge\omega$
is the curvature two - form of the local $SO(n-1,1)$ gauge invariance. In
component form it is given by equation (17).

In order to show the equivalence of (21) with ordinary general relativity, we
must demand $\det(e_{i}^{a})\neq 0$. Since we can make sense of the action (21)
and hence the field equations even when $\det(e_{i}^{a})=0$, it is clear
the action (21) represents an extension of ordinary general relativity
\cite{horo2}. In any case, we will take the action (21) as our action for
gravity in $n$ space - time dimensions ($n>2$).

\section{First order quantum gravity}
Our starting point is pure gravity in $n$ - space - time dimensions ($n>2$)
in the first order formalism with zero cosmological constant.
The classical action of interest is thus:
\begin{equation}
S=\int_{M_{n}}e^{a}\wedge e^{b}\wedge...\wedge R^{cd}(\omega)\epsilon_{ab...cd},
\end{equation}
where $M_{n}$ is some fixed manifold. We could add to this action other
terms such as the square of the curvature tensor and higher derivative terms
etc. which are consistent with the symmetries of the theory.
Under these circumstances the field content of the classical theory will
remain
the same but there may be additional ghost fields to worry about which are
not part of the standard gauge fixing prescription as presented below. We
therefore consider the classical action as given by equation (22).

We now wish to regard the fields $e$
and $\omega$ as gauge fields unrelated to any space - time metrics. We therefore
regard $e$ and $\omega$ to be propagating on $M_{n}$ and since the action
(22) does not contain any metric of $M_{n}$ we can view this action
as a Schwarz type topological field theory and thus the full
machinery of section (2) can be exploited.

Due to the local gauge invariance we must gauge fix in order to quantise the
theory. As in section (2), we must pick some metric $g_{ij}$ on
$M_{n}$, unrelated to the one - forms $e_{i}^{a}$, to impose some gauge fixing
condition. We choose the Landau gauge:
\begin{equation}
D^{i}\omega^{ab}_{i}=D^{i}e^{a}_{i}=0,
\end{equation}
where $D^{i}$ is the gravitational covariant derivative with respect to the
metric $g_{ij}$. It is now convenient to construct the relevant BRST variations
$\delta_{Q}$ of the various fields in our problem. From equations (18) and
(19), they are given by:
\begin{equation}
\delta_{Q}e^{a}_{i}=-\partial_{i}c^{a}-\omega^{a}_{i\; b}c^{b}+e_{id}
f^{ad}=-(\tilde{D}_{i}c)^{a}+e_{id}f^{ad},
\end{equation}
\begin{equation}
\delta_{Q}\omega_{i}^{ab}=-\partial_{i}f^{ab}-\omega_{ci}^{\;\; b}f^{ac}
+\omega^{a}_{i\;\; c}f^{cb}=-(\tilde{D}_{i}f)^{ab},
\end{equation}
where $c^{a}$ and $f^{ab}$ are respectively the ghosts associated with the
gauge invariance of $e^{a}_{i}$ and $\omega^{ab}_{i}$.
Recalling we may enforce
our gauge conditions (23) using Lagrange multipliers $u^{a}$ and $v^{ab}$
such that:
\begin{equation}
\delta_{Q}\bar{c}^{a}=u^{a},\;\; \delta_{Q}u^{a}=0,
\end{equation}
and:
\begin{equation}
\delta_{Q}\bar{f}^{ab}=v^{ab},\;\; \delta_{Q}v^{ab}=0,
\end{equation}
where $\bar{c}^{a}$ and $\bar{f}^{ab}$ are anti - ghosts, the ghost plus gauge
fixing terms
$S_{gf}$ corresponding to the classical action $S$ is therefore given by:
\begin{equation}
S_{gf}=\delta_{Q}\int_{M_{n}}\sqrt{g}(\bar{c}_{a}D^{i}e_{i}^{a}+\bar{f}_{ab}
D^{i}\omega_{i}^{ab}).
\end{equation}
Evaluating we obtain:
\begin{equation}
S_{gf}=\int_{M_{n}}\sqrt{g}(g^{ij}u_{a}D_{i}e^{a}_{j}+g^{ij}v_{ab}D_{i}
\omega^{ab}_{j}
-g^{ij}\bar{c}_{a}\tilde{D}_{i}D_{j}c^{a}-g^{ij}\bar{f}_{ab}\tilde{D}_{i}D_{j}
f^{ab}+\bar{c}_{a}e_{id}D^{i}f^{ab}).
\end{equation}
Note, the last term in equation (29) may be ignored since if in $S_{gf}$ we
integrate over the $f^{ab}$ fields we will obtain the delta functional
constraint
$\delta(-g^{ij}\bar{f}_{ab}\tilde{D}_{i}D_{j}+\bar{c}_{a}e_{ib}D^{i})$. When
substituted back into $S_{gf}$ we clearly reproduce the fourth term in $S_{gf}$.

The full partition function for pure gravity in the first order formalism
is then formally given by:
\begin{equation}
Z(M_{n},g_{ij})=\int DeD\omega DuDvD\bar{f}DfD\bar{c}Dc\exp i(S+S_{gf}).
\end{equation}
We stress at this point we are not worried about perturbative 
renormalisability. Our interest is in the full non - perturbative partition
function which we assume makes sense.

Since the BRST variations do not involve the metric $g_{ij}$, it is clear
the demonstration of topological invariance given in 
section (2) for topological field theories also applies to our pure
gravitational
case with the partition function given by (30). That is, the classical action
(22) does not contain the gauge fixing metric $g_{ij}$ and therefore the
partition function (30) will be a topological invariant with respect to the
gauge fixing metric $g_{ij}$. It then follows that in order
to ensure the absence of any topological anomalies with respect to
$g_{ij}$, we must demand the
index $K$ given by equation (12) vanishes. We check this as follows.

The full particle content involved in evaluating the partition function (30)
is:
\begin{equation}
\{ e^{a}_{i},\; \omega^{ab}_{i},\; u_{a},\; v_{ab},\; f_{ab},\; \bar{f}_{ab},\;
c_{a},\; \bar{c}_{a}\}.
\end{equation}
Since we have to ensure the full gauged fixed partition function is invariant
under BRS - transformations, we must insert an appropriate factor of
$e=\det(e^{a}_{i})$ \cite{FY}. That is, the appropriate BRST invariant and
hence gauge independent combination
of fields which needs to appear in the measure of the partition function (30)
is thus:
\begin{equation}
\{ e^{L}e^{a}_{i},\; \omega^{ab}_{i},\; u_{a},\; v_{ab},\; f_{ab},\;
\bar{f}_{ab},\; c_{a},\; \bar{c}_{a}\},
\end{equation}
where $L$ is an appropriate factor which will ensure BRST invariance of the
path integral measure and is given by (remembering to keep track of the
tensor nature and statistics of the fields involved):
\begin{equation}
L=2\times n(n\frac{(n-2)}{4n}-\frac{1}{4})+\frac{n(n-1)}{2}(n\frac{n-2}{4n}
-\frac{1}{4})=2\times \frac{n(n+1)}{2}\frac{(n-3)}{4},
\end{equation}
where the factor of 2 is due to the vielbein being related to the ^^ ^^ square"
of its corresponding metric tensor. The partition function which will now
be invariant under BRS - transformations is therefore given by:
\begin{equation}
Z(M_{n},g_{ij})=\int DeD\omega DuDvD\bar{f}DfDcD\bar{c}.e^{L}.\exp i(S+S_{gf}).
\end{equation}
The index of equation (10) associated with this particle content is therefore
given by:
\begin{equation}
K=Ln(\frac{n(n-2)}{4n})+L=L[\frac{n^{2}-2n+4}{4}]=2\times \frac{n(n+1)}{2}
(\frac{n-3}{4})[\frac{n^{2}-2n+4}{4}].
\end{equation}
Clearly $K=0$ only for $n=3$. This of course corresponding to the computable
Chern - Simons - Witten topological gravity \cite{Wit2,Wit3}. Thus, we have
confirmed that in $n=3$ dimensions, first order quantum gravity has an
interpretation as a topological field theory with respect to some gauge
fixing metric $g_{ij}$.

The constraint to $n=3$ would seem to suggest that our approach is totally
useless in attempting to predict any useful physics. However, the Universe
we see today is not that of pure gravity but one including matter and gauge
fields. Our next task, therefore, is to extend the pure gravitational case to
include matter and gauge fields.

Consider then $s$ complex scalar fields $\phi_{p}$ ($p$ = 1, 2, ..., $s$)
together
with $f$ fermionic fields $\psi_{q}$ ($q$ = 1, 2, ..., $f$) with each
field transforming under some representation of a Yang - Mills
gauge group $G$. These fields will live in the physical relevant  space
described by
the vielbein $e$ and spin connection $\omega$ and thus the full classical
action of the combined fields is now given by:
\begin{equation}
S=\int_{M_{n}} e\wedge e... \wedge R(\omega)+S_{gauge}+S_{\phi}+S_{\psi},
\end{equation}
where $S_{gauge}$ is the kinetic term associated with the Yang - Mills
gauge group $G$:
\begin{equation}
S_{gauge}=\int\det(e)F^{ij}F_{ij},
\end{equation}
where $F_{ij}$ is the covariant field strength associated with the gauge
group $G$. $S_{\phi}$ is the classical action for the complex scalar fields:
\begin{equation}
S_{\phi}=\int\det(e)\eta^{ab}e_{a}^{i}e_{b}^{j}D_{i}\phi^{*}
D_{j}\phi+...,
\end{equation}
where $D_{i}$ is the covariant derivative with respect to the Yang - Mills
gauge group $G$
and the dots represent possible gauge invariant mass and interaction terms.
Finally, $S_{\psi}$ is the classical action of the fermion fields:
\begin{equation}
S_{\psi}=\int\det(e)e^{i}_{a}\bar{\psi}\gamma^{a}D_{i}\psi+...,
\end{equation}
where again the dots represent possible gauge invariant mass and interaction
terms and here $D_{i}$ is the covariant derivative with respect to the spin
connection $\omega$ and Yang - Mills gauge group $G$. Clearly, if we are
interested in chiral fermions we must insist $n$ be even.

Note that we can only write these classical actions down on classical
manifolds for which $\det(e)\neq 0$ and spin structures exist etc. One
of the main motivations for this work is to try and understand how the
quantum theory takes care of these fields when one attempts to integrate
over all metrics. These fields must in some sense melt away on non -
classical manifolds since we can only make sense of them in physical
acceptable spaces. We will nonetheless continue with the hope that some
light can be shed on this problem.

To quantise the theory we must again gauge fix. We do this as in the case of
pure
gravity but we must also, of course, gauge fix the gauge field associated
with the gauge group $G$. This Yang - Mills gauge fixing being performed in
the physically relevant space defined by the vielbein $e_{i}^{a}$.

The demonstration of topological invariance of
section (2) is unaffected by the inclusion of our fields provided the index
$K$ is zero. Again, however, we must insert an appropriate factor $e^{L}$
into the
corresponding partition function in order to ensure BRST invariance with
respect to all the fields of interest. Our combination of fields now gives:
\begin{equation}
L=2\times[\frac{n(n+1)}{2}\frac{(n-3)}{4}+\dim_{adj}(G)\frac{(n-3)}{4}+
\frac{2}{4}\sum_{p=1}^{s}\dim(\phi_{p})-\frac{2}{4}\sum_{q=1}^{f}
\dim(\psi_{q})],
\end{equation}
where $\dim_{adj}(G)$ is the dimension of the adjoint representation of the
gauge group $G$, $\dim(\phi_{p})$ and $\dim(\psi_{q})$ are the dimensions of
the possible representations the scalar and fermion fields may transform
under the gauge group $G$ respectively, the factor of 2 in front of the
summations are because the scalar and fermion fields are complex since we
wish to couple these fields to gauge fields.
The factor of 2 in front of the square bracket is again due to some metric
tensor being the ^^ ^^ square" of the vielbein.

With this appropriate factor, our full particle content for the theory is:
\begin{equation}
\{e^{L}e^{a}_{i},\; \omega^{ab}_{i},,\; u_{a},\; v_{ab},\; f_{a},,\;
\bar{f}_{a},\; c_{ab},\; \bar{c}_{ab},\; A_{i},\; b,\; h,\; \bar{h},\;
\phi,\; \psi\},
\end{equation}
where $A_{i}$ is the gauge field associated with the Yang - Mills gauge
group $G$ (gauge
group indices not shown) and $b$, $h$ and $\bar{h}$ are the corresponding
Lagrange multiplier field (which enforces some Yang Mills gauge constraint),
ghost
and anti - ghost fields respectively. We may now formally construct the
partition function
for this particle content and calculate the index $K$.  We therefore
deduce:

For topological invariance to be preserved in the full quantum
theory with respect to the gauge fixing metric $g_{ij}$, we must demand:
$$
K=2\times[\frac{n(n+1)}{2}\frac{(n-3)}{4}+\frac{(n-3)}{4}\dim_{adj}(G)+
\frac{2}{4}\sum_{p=1}^{s}\dim(\phi_{p})-\frac{2}{4}\sum_{q=1}^{f}\dim(\psi_{q})]
$$
\begin{equation}
\end{equation}
$$\times[\frac{n^{2}-2n+4}{4}]=0.$$
Inspection of
equation (42) shows a kind of ^^ ^^ supersymmetry" at work here in the
sense the contribution of all the fermion fields must be canceled by the
contribution of all the integer spin fields in a given space - time dimension.
This presumable is tied up with the fact that topological field theories have
no degrees of freedom on the topological space under consideration. In our
case, the space associated with the gauge fixing metric $g_{ij}$ and not the
physically interesting space defined via the vielbein fields $e_{i}^{a}$, which
we now discuss.
\section{Degrees of freedom}
It is well known that topological quantum field theories have no physical
degrees of freedom. This is because every field in the theory has an associated
^^ ^^ ghost" field which is identical in every way except it has opposite
statistics \cite{Thom}. If, then, we are to regard first order quantum
gravity as a topological field theory we must show it has no physical degrees
of freedom on an appropriate space. In our case, the topological space being
the gauge fixing metric $g_{ij}$.

It is well known that gauge fields in an $n$ dimensional space - time has
$n-2$ physical degrees of freedom. Since in our case we have gauge fields
transforming under different groups (Poincar$\acute{e}$ and Yang - Mills
groups) we need to take gauge degrees of freedom into account when calculating
total degrees of freedom. One might therefore guess that if we
have a gauge field of a gauge group $G$ in an $n$ dimensional space - time
, the total number of space - time plus gauge degrees of freedom will be
$(n-2)\times \dim_{adj}(G)$. This is, however, incorrect. We see this as
follows.

One way of counting degrees of freedom is to count the number of Laplacian
determinants which appear in partition functions from quadratic operators.
A bosonic like determinant $(\det\Delta)^{-1/2}$ contributing +1 and a
fermionic like determinant $(\det\Delta)^{1/2}$ contributing $-1$ local
bosonic degrees of freedom.
Consider then a pure gauge field in an n dimensional space - time with field
strength $F_{ij}$. The classical action expanded to quadratic order is given
by:
\begin{equation}
\int F^{a}_{ij}F^{ij}_{a}\approx\int A_{i}^{a}(\partial_{k}\partial^{k}\eta
^{ij}-\partial^{i}\partial^{j})A_{ja}=\int A_{i}^{a}\Delta^{ij}A_{aj}.
\end{equation}
The fully gauged fixed partition function (to second order) is thus given by:
\begin{equation}
Z_{2}=\int DA_{i}^{a}Du_{a}Dc_{a}D\bar{c}^{a}\exp i\int(A_{i}^{a}\Delta^{ij}
A_{ja}+u_{a}\partial^{i}A_{i}^{a}+\bar{c}_{a}\partial^{i}D_{i}c^{a}),
\end{equation}
where $u_{a}$ is a Lagrange multiplier which enforces the gauge constraint
$\partial^{i}A^{a}_{i}=0$ (here we work in flat $n$ dimensional space - time),
$c^{a}$ and $\bar{c}_{a}$ are the ghost and anti - ghost fields, $i=0,1,...,
n-1$ and $a=1,2,...,\dim_{adj}(G)$. Integrating
over the gauge and ghost fields gives (ignoring zero modes):
\begin{equation}
Z_{2}\sim\int Du_{a}\frac{\prod_{a}\det(\partial^{i}D_{i})}{\prod_{i,a}(\det
\Delta)^{1/2}}\exp i\int \partial^{i}u_{a}\frac{1}{\Delta^{ij}}\partial
^{j}u_{a}.
\end{equation}
If we were only interested in space - time degrees of freedom we would stop
here since the Lagrange multiplier field $u_{a}$ is non - propagating.
Counting degrees of freedom we have $n$ bosonic and 2 fermionic
determinants giving a total of $n-2$ degrees of freedom. Since we also need
to count gauge degrees of freedom, we must now perform the $u_{a}$ integral:
\begin{equation}
Z_{2}\sim\frac{\prod_{a}\det(\partial^{i}D_{i})}{\prod_{i,a}(\det\Delta
)^{1/2}}\times \prod_{a}(\det(\frac{\Delta}{\partial\partial}))^{1/2}.
\end{equation}
We therefore see the total number of bosonic degrees of freedom for a single
gauge boson in an $n$ dimensional space - time is:
\begin{equation}
(n-2)\dim_{adj}(G)-\dim_{adj}(G)=(n-3)\dim_{adj}(G).
\end{equation}
Similarly, after linearising the curvature two form $R$ of equation (21) and
performing the integrals over the vielbein and spin connection fields (up the
quadratic order) simultaneously, we find the vielbein and spin connection
fields contributes $\frac{n(n+1)}{2}(n-3)$ bosonic degrees of freedom.

Returning back to our particle content of (41), we now see the total number of
local bosonic degrees of freedom $D_{g}$ is given by:
\begin{equation}
D_{g}=\frac{n(n+1)}{2}(n-3)+(n-3)\dim_{adj}(G)+2\sum_{p=1}^{s}\dim(\phi_{p})
-2\sum_{q=1}^{f}\dim(\psi_{q}),
\end{equation}
where the subscript $g$ refers to the topological space of interest.
We therefore deduce:
\begin{equation}
K=2\times\frac{D_{g}}{4}\times[\frac{n^{2}-2n+4}{4}].
\end{equation}
Clearly $K=0$ when $D_{g}=0$ and thus quantum gravity, when coupled to other
fields, will have no local bosonic degrees of freedom propagating on the gauge
fixing metric $g_{ij}$ and therefore  strengthening its interpretation as a
topological field theory.

We finally note the real physical bosonic degrees of freedom $D_{real}$
propagate on the space defined via the vielbein field $e$. For fixed $e$ and
spin connection $\omega$, it is given by:
\begin{equation}
D_{real}=(n-3)\dim_{adj}(G)+2\sum_{p=1}^{s}\dim(\phi_{p})-2\sum_{q=1}^{f}
\dim(\psi_{q}).
\end{equation}

\section{Particle physics}
Let us now consider the vanishing of the index $K$ in the context of particle
physics. Let us further restrict ourselves to the physically interesting ball
park of $n=4$. Equation (42) then becomes:
\begin{equation}
K=\frac{6}{4}[10+\dim_{adj}(G)+2\sum_{s}\dim(\phi_{p})-2\sum_{f}\dim(\psi_{f})]
=0.
\end{equation}
The first thing we notice is our constraint implies the dimension of the
gauge group $G$ must be even. Since $\dim(Su(3)\otimes Su(2)\otimes U(1))=12$
our first instinct is to see if the standard model of particle physics
respects the constraint. The quantum numbers of a single family with respect
to $SU(3)\otimes SU(2)\otimes U(1)$ are:
\begin{equation}
(3,2)_{1/6}\oplus (\bar{3},1)_{-2/3}\oplus(\bar{3},1)_{1/3}\oplus(1,2)_{-1/2}
\oplus(1,1)_{1},
\end{equation}
where the first entry in the parenthesis indicates the representation under
$SU(3)$, the second one of SU(2) and the subindex the weak hypercharge $Y$.
There is also a doublet Higgs transforming as $(1,2)_{1/2}$.

It is easily shown that the particle content of the standard model does not
give zero $K$. In fact, one obtains $K=\frac{6}{4}(-64)$. One can now
force $K$ to be zero by adding appropriate matter fields to the standard
model. For example, an appropriate set
of fields which will cancel the $-64$ in our constraint is to add a $SU(5)$
gauge theory together with two complex scalar fields transforming as a
$\bf{10}$ under $SU(5)$. The particles of the standard model are then required
to transform as singlets under this $SU(5)$. Thus, with this particle content,
we can interpret [standard model]$\otimes[\bf{24}\oplus \bf{10}\oplus\bf{10}]$,
where the second bracket refers to transformations under $SU(5)$ of gauge
bosons and complex scalar fields,
as a topological field theory, with respect to the gauge fixing metric,
when coupled to gravity.

As in
string theory, we may interpret this unseen bosonic matter as hidden or dark
matter.
There is clearly no mixing between this matter and the standard model and we
may therefore argue that this hidden sector interacts with our universe
gravitationally only.

Similar remark applies to the unification gauge group
$SU(5)$ with the three families in reducible representations
$3(\bar{\bf{5}}+\bf{10})$ together with two Higgs fields transforming as
$\bf{24}$ and $\bf{5}$. With this particle content we obtain $K=\frac{6}{4}
(2)$. In this case, we can easily force $K$ to be zero by considering a slight
extension of the usual $SU(5)$ model. We simply add a singlet complex
fermion field
(one not three) which we may interpret as a right - handed neutrino.  So,
we can couple the standard $SU(5)$ model plus a right - handed
singlet neutrino to gravity and interpret this theory as a topological field
theory with respect to the gauge fixing metric.

Of course, we can go on in this manner and classify all possible interesting
theories which respects the constraint (43). By inspection, one can guess we
can generate a class of models similar to what one sees in string theory.
However, our models will have no ugly mixing between observed and so called
hidden sectors and gauge groups can easily be constructed to have low rank.

Let us  again remark on two important observations which are manifest
from the constraint (51) in $n=4$ space - time dimensions. First, the adjoint
representation of gauge fields must add up to an even number and,
as in an arbitrary $n$ dimensional space - time, there is a
kind of ^^ ^^ supersymmetry" at work in the sense the half spin fields
must cancel out all contributions of the integer spin fields. Let us finally
return to this observation in the context of $n$ - dimensional space - time.

If one believes in unification, one must believe in supersymmetry since the
unification of coupling constants is most naturally achieved in
supersymmetric field theories around $10^{16}$ Gev. In particular, $N=1$
space - time supersymmetry
is needed if one desires chiral fermions. As we are discussing gravity it
seems most
natural to extend our ideas to supergravity. However, this program is not so
straight forward in an arbitrary n - dimensional space - time since, for
example, $N=1$ space - time supersymmetry cannot be realised in arbitrary
space - time dimensions if one demands particle helicities not to exceed
2 \cite{stra}.

The supersymmetric like generalisation we have in mind is in the following
sense.
We have been viewing gravity as a gauge theory of the Poincar$\acute{e}$ group
with vielbein fields $e_{i}^{a}$ and spin connection fields $\omega_{i}^{ab}$.
Since these fields are spin - 1 vector bosons we can write down a 
supersymmetric like version by adding to the theory fermionic partner fields.
That is, add two adjoint real spin 1/2 fermions with one transforming as
$e_{i}^{a}$ and the other as $\omega_{i}^{ab}$ and similarly for the vector
fields of the Yang - Mills gauge group $G$.
Likewise, all matter fields we wish to couple to the theory are accompanied
with partners of opposite statistics. A complex scalar field is added with a
Weyl, or some
complex fermion if $n$ is not even, transforming identically as the complex
scalar field under the Yang - Mills gauge group $G$, and vice versa. Note we
are not worrying about equal number of fermion and boson states in a given
multiplet here.
With this set up, we
see the matter fields will not contribute to the index $K$, the only
contribution coming from all adjoint vector and fermion fields of the
Poincar$\acute{e}$ group and Yang - Mills gauge group $G$. With this
supersymmetric like model, we deduce:
$$
K_{super}=2\times[\frac{n(n+1)}{2}(\frac{n-3}{4}-\frac{1}{4})+(\frac{n-3}{4}
-\frac{1}{4})\dim_{adj}(G)]\times[\frac{n^{2}-2n+4}{4}]$$
\begin{equation}
=2\times[\frac{n(n+1)}{2}\frac{(n-4)}{4}+\frac{(n-4)}{4}\dim_{adj}(G)]\times
[\frac{n^{2}-2n+4}{4}],
\end{equation}
where we note a real fermion contributes $-1/4$ to $K_{super}$ in a given
gauge representation (a complex fermion contributing $-2/4$).
Clearly, for $K_{super}=0$ we must demand $n=4$. This being the first time a
theory has picked out the correct number of space - time dimensions!
Unfortunately, this approach will tell us nothing regarding possible matter
or Yang - Mills gauge content of the theory.

\section{Conclusion}
In this paper, we have attempted to give quantum gravity, in its first order
formalism, an interpretation as a topological field theory with respect to
some gauge fixing metric $g_{ij}$. We coupled gravity to gauge, scalar and
fermion fields and argued the theory resembles a Schwartz type topological
field theory if we regard all the fields in the classical action to be
independent of $g_{ij}$.

By examining the path integral measure
associated with the partition function, we discovered a constraint on the
particle content of the theory together with the dimension of possible gauge
groups and the number of space - time dimensions. Respecting this constraint
can be thought of as a cancellation of a topological anomaly. Finally, we
discussed possible models which respected this constraint.

One of the major issues confronting any formulation of non - perturbative
quantum gravity is constructing the theory independently of any background
metric \cite{Ash}. One would like space - time to emerge from the theory without
artificially introducing a given metric in order to perform calculations.
Perhaps a nice way to by pass these difficulties is via the construction
discussed in this paper. One can choose one's favorite space - time metric as
a gauge fixing metric, perform calculations and show that your answers does
not depend on the choice of your space - time metric. This gauge fixing metric
clearly is not a physical metric with physical degrees of freedom
propagating. Its primary purpose is to provide an arena for us to perform
calculations in a generally covariant way.

Traditionally, the cancellation of anomalies is our root understanding of
certain physical classifications, such as the particle content in a single
family of the standard model where we are forced to arrange fermions into
gauge anomaly free representations, as well as physical processes
involving certain decay channels.
The more constraints via anomalies one can impose on a
theory the better chance we have of understanding its degree of uniqueness.
Together with gauge and gravitational anomalies, topological anomalies may
help us write down an unique theory of our universe.

\end{document}